\documentclass[manuscript,screen]{acmart}
\usepackage[table]{xcolor}
\usepackage{tabularx}
\usepackage{subcaption}
\usepackage{enumitem}
\definecolor{LabelColor}{HTML}{004E8A}
\newcommand{\LabelText}[1]{\textcolor{LabelColor}{#1}}
\AtBeginDocument{%
  }

\setcopyright{acmlicensed}
\copyrightyear{2026}
\acmYear{2026}
\setcopyright{cc}
\setcctype{by}
\acmConference[LAK 2026]{LAK26: 16th International Learning Analytics and Knowledge Conference}{April 27-May 01, 2026}{Bergen, Norway}
\acmBooktitle{LAK26: 16th International Learning Analytics and Knowledge Conference (LAK 2026), April 27-May 01, 2026, Bergen, Norway}
\acmPrice{}
\acmDOI{10.1145/3785022.3785088}
\acmISBN{979-8-4007-2066-6/2026/04}

\begin{document}

\title{How to Assess AI Literacy: Misalignment Between Self-Reported and Objective-Based Measures} 









\author{Shan Zhang}
\authornote{Both authors contributed equally to this research.}
\orcid{0009-0003-3532-0661}
\email{zhangshan@ufl.edu}
\affiliation{%
  \institution{University of Florida}
  \city{Gainesville}
  \state{FL}
  \country{USA}
}

\author{Ruiwei Xiao}
\authornotemark[1]
\orcid{0000-0002-6461-7611}
\email{ruiweix@andrew.cmu.edu}
\affiliation{%
  \institution{Carnegie Mellon University}
  \city{Pittsburgh}
  \state{PA}
  \country{USA}
}

\author{Anthony F. Botelho}
\email{abotelho@coe.ufl.edu}
\orcid{0000-0002-7373-4959}
\affiliation{
  \institution{University of Florida}
  \city{Gainesville}
  \state{FL}
  \country{USA}
}

\author{Guanze Liao}
\email{gzliao@mx.nthu.edu.tw}
\orcid{0000-0002-5758-3396}
\affiliation{
  \institution{National Tsing Hua University}
  \city{Hsinchu}
  \country{Taiwan}
}

\author{Thomas K. F. Chiu}
\email{tchiu@cuhk.edu.hk}
\orcid{0000-0003-2887-5477}
\affiliation{
  \institution{The Chinese University of Hong Kong}
  \city{Hong Kong}
  \country{Hong Kong}
}

\author{John Stamper}
\email{jstamper@cmu.edu}
\orcid{0000-0002-2291-1468}
\affiliation{
  \institution{Carnegie Mellon University}
  \city{Pittsburgh}
  \state{PA}
  \country{USA}
}

\author{Kenneth R. Koedinger}
\email{kk1u@andrew.cmu.edu}
\orcid{0000-0002-5850-4768}
\affiliation{
  \institution{Carnegie Mellon University}
  \city{Pittsburgh}
  \state{PA}
  \country{USA}
}

\renewcommand{\shortauthors}{Shan Zhang et al.}

\begin{abstract}
The widespread adoption of Artificial Intelligence (AI) in K–12 education highlights the need for psychometrically-tested measures of teachers' AI literacy. Existing work has primarily relied on either self-report (SR) or objective-based (OB) assessments, with few studies aligning the two within a shared framework to compare perceived versus demonstrated competencies or examine how prior AI literacy experience shapes this relationship. This gap limits the scalability of learning analytics and the development of learner profile–driven instructional design. In this study, we developed and evaluated SR and OB measures of teacher AI literacy within the established framework of Concept, Use, Evaluate, and Ethics. Confirmatory factor analyses support construct validity with good reliability and acceptable fit. Results reveal a low correlation between SR and OB factors. Latent profile analysis identified six distinct profiles, including overestimation (SR > OB), underestimation (SR < OB), alignment (SR $\approx$ OB), and a unique low-SR/low-OB profile among teachers without AI literacy experience. Theoretically, this work extends existing AI literacy frameworks by validating SR and OB measures on shared dimensions. Practically, the instruments function as diagnostic tools for professional development, supporting AI-informed decisions (e.g., growth monitoring, needs profiling) and enabling scalable learning analytics interventions tailored to teacher subgroups.

\end{abstract}



\maketitle

\section{Introduction} 

The rapid integration of Artificial Intelligence (AI) into K–12 education has driven shifts in many aspects, including policy, curriculum, and teacher professional development. At the policy level, both international and national organizations now view AI competency as essential for future readiness. UNESCO, for example, has released global guidance on AI in education \cite{unesco_ai_cft_2024,unesco2023genai}, while governments have introduced initiatives such as the \textit{Executive Order on AI Education} in the U.S. \cite{whitehouse2025ai}, the EU’s \textit{Digital Education Action Plan} \cite{eu2021digital}, and China's \textit{Education Modernization Plan 2024–2035} \cite{china2025edustrategy}. In response, education systems are introducing AI-focused curricula and expanding professional development for teachers. Taiwan, for instance, requires AI instruction starting in middle school under its centralized \textit{108 Curriculum Guidelines}, which frame AI competencies, covering concepts, applications, and ethics, as cross-cutting learning goals \cite{naer108curriculum}. These efforts point to an urgent need to prepare educators with knowledge and skills to teach, use, and integrate AI responsibly \cite{chiu2025responsible, wang2023k,song2024framework}.

In response to these needs, researchers have proposed numerous AI literacy frameworks for educators, spanning conceptual, technical, and ethical competencies \cite{ng2021conceptualizing,ning2024teachers}, and professional development (PD) programs have been co-designed to upskill teachers accordingly \cite{hutchins2025empowering}. For instance, \citet{ng2021conceptualizing} articulated four dimensions of an AI literacy framework: Know \& Understand, Use \& Apply, Evaluate \& Create, and Ethics. Similarly, \citet{long2020ai} provided a comprehensive definition of AI literacy, outlining 17 competencies and 15 design considerations (e.g., core AI concepts and processes, human--AI interaction, and societal/ethical impacts) to guide curriculum and assessment design. In parallel, both national \cite{iste_key_initiatives} and international \cite{figueiredo2025ai_literacy_programs,TeachAI_Toolkit} initiatives provide support to educators through AI policies, toolkits \cite{TeachAI_Toolkit}, self-paced MOOCs \cite{google-ai-educators}, and micro-credential training programs \cite{hutchins2025empowering}.

Beyond frameworks and PD, both self-report (SR) and objective-based (OB) assessments have been developed to measure teachers' perceptions and demonstrated knowledge \cite{yue2024understanding,polak2022teachers,jin2025glat}. For example, \citet{yue2024understanding} implemented SR questions and surveyed 1,831 K–12 teachers on Technological Pedagogical Content Knowledge (TPACK) readiness and attitudes toward AI education, exemplifying scalable instruments that capture perceptions and dispositions of teachers' AI literacy; \citet{jin2025glat} measured factual and applied AI literacy skills across four dimensions using OB instrument through the Generative AI Literacy Test.

Despite a growing body of SR and OB measures, few studies have examined the relationship between these two types of instruments. In particular, it remains unclear whether teachers' perceptions of their AI literacy align with their demonstrated competencies, or how this relationship varies based on their prior AI literacy learning experience. Investigating this underexplored alignment between SR and OB responses is essential for learning analytics (LA) in the AI literacy context: it enlights learning designers' choice of instruments (e.g., when to use SR and/or OB), enables more accurate diagnosis of teachers’ proficiency, and thereby supports more targeted interventions/PD for distinct subgroups. Insights from this alignment can also strengthen theoretical models and inform evidence-based instructional design to better prepare educators to integrate AI effectively and responsibly, ultimately fostering a more AI-literate teaching workforce. To address these issues, we pose the following research questions:


\begin{enumerate}[label=, leftmargin=10pt, labelsep=0pt, itemsep=2pt, topsep=2pt, parsep=0pt]
    \item \textbf{RQ1}: To what extent does the objective measure-based AI literacy assessment demonstrate psychometric stability?
    \item \textbf{RQ2}: What factors emerge from K–12 teachers' self-reported and objectively measured levels of AI literacy?
    \item \textbf{RQ3}: What distinct learner profiles emerge from the self-report- and objective-based factors?
    \item \textbf{RQ4}: How do these profiles differ between teachers with prior AI literacy experience and those without?
\end{enumerate}

\section{Related Work} 
\subsection{AI Literacy For K-12 Educators}
There is a growing consensus that AI literacy is essential for all learners \cite{wang2023k, song2024framework}, which has led to notable efforts to better define, characterize, and measure the construct, particularly in relation to how individuals engage with AI. According to \citeauthor{ng2021conceptualizing}'s \cite{ng2021conceptualizing} review of 30 peer-reviewed studies, AI literacy encompasses basic AI knowledge and abilities for working with AI, motivation for learners' future careers, and an understanding of ethical concerns necessary to use AI responsibly. AI literacy instruction is especially critical in K--12 education, where teachers play a central role. Specifically, teachers’ AI literacy is positioned as a critical prerequisite for designing and implementing such instruction \cite{nazaretsky2022teachers}. In other words, educators must first demonstrate AI literacy by adapting to teaching and working in AI-integrated environments before they can confidently guide their students to engage with AI effectively and responsibly \cite{ng2023teachers}. 

To enhance educators' AI literacy, researchers have developed frameworks, resources, and evaluative measures for teacher education \cite{mikeladze2024comprehensive,black2024framework}. For example, \citet{vyortkina2024ai} proposed a scalable and sustainable professional-development model that specifies guiding principles, modular content domains, and criteria for tool selection and iterative evaluation to build teachers' AI literacy. Likewise, \citet{chiu2021holistic} identified six key components of an AI K--12 curriculum through individual interviews, teaching documents, and meetings with 24 teachers. Beyond these teacher-oriented approaches, other efforts have integrated AI literacy into established educational theories and digital literacy models, extending them to today's AI-enhanced learning environments. For instance, AI-TPACK extends the TPACK framework by embedding AI into teachers' technological, pedagogical, and content knowledge \cite{ning2024teachers}, while \citet{ng2022ai} incorporated AI literacy into Bloom's Taxonomy. Collectively, these initiatives advance a more systematic foundation for supporting educators' AI literacy and its integration into teaching and learning.

\subsection{Professional Development for AI Literacy in K–12}


With the growing importance of upskilling teachers for AI-integrated education, PD programs and micro-credentials have been developed by researchers and NGOs to translate theory into practice based on existing frameworks \cite{zhang2025learning}. For example, \citet{hutchins2025empowering} co-designed an AI microcredential with K--12 educators using conjecture mapping and memoing across three workshops to identify essential themes (e.g., teaching ethical AI in K--12) and requirements (e.g., quick, easily accessible, asynchronous learning activities) for effective teacher PD. Likewise, \citet{wu2025enhancing} implemented a two-day AI-TPACK workshop with 25 elementary teachers in Taiwan, resulting in significant gains in AI competencies across all targeted constructs (AI Knowledge, Application, Integration, and Ethical Considerations in Teaching). At a broader scale, the \textit{Day of AI} initiative by MIT RAISE provides free, research-based curricula and twice-weekly training (in partnership with i2Learning \cite{i2learning}) to support teachers in integrating AI literacy into existing curricula \cite{breazeal2023day}. Similarly, Code.org has expanded its long-standing CS education initiatives to include AI-focused teacher training, offering workshops and self-paced learning programs that have reached millions of educators worldwide \cite{codeorgTeachers2025}. Meanwhile, \textit{TeachAI}, a global coalition of education organizations and companies, provides the \textit{AI Guidance for Schools Toolkit} to help education authorities, school leaders, and teachers develop responsible, context-sensitive guidance for integrating AI into education \cite{TeachAI_Toolkit}. Collectively, these initiatives highlight the growing global efforts to equip K--12 educators with the knowledge and skills needed to responsibly integrate AI into teaching and learning.

\subsection{Measuring Teachers' AI Literacy using Both Self-Report- and Objective Measure-Based Assessment}

Given the theoretical frameworks and ongoing efforts to provide teacher training, measuring the effectiveness of these initiatives remains a challenge. Researchers have used various measures to assess teachers’ AI literacy competencies and the learning gains from related learning activities. Most existing studies rely on self-reported measures, capturing dimensions such as teachers’ readiness, attitudes, and intentions to teach AI \cite{yue2024understanding,polak2022teachers,ning2025development,tenberga2024artificial,younis2025artificial,carolus2023mails}. For example, \citet{polak2022teachers} surveyed 135 teachers to examine self-reported digital competence; \citet{yue2024understanding} assessed 1,831 K–12 teachers on TPACK readiness and attitudes toward AI education. Extending this work beyond teacher-specific instruments, \citet{carolus2023mails} applied AI literacy scales originally designed for the general public, measuring psychological competencies alongside AI literacy to explore associations with longer-term AI use. Similarly, \citet{laupichler2023development} created a 31-item AI literacy scale for non-experts, encompassing Technical Understanding, Critical Appraisal, and Practical Application, and validated its content through an iterative Delphi study with 53 subject-matter experts \cite{laupichler2023development,laupichler2023delphi}.

While self-reported measures are useful for capturing teachers’ perceptions and attitudes, they may not accurately reflect actual competencies \cite{keefer2015self,xiao2025improving,chiu2024developing}. To address this limitation, researchers have developed objective, performance-based assessments, though most have been psychometrically tested with students rather than educators \cite{jin2025glat,chai2021perceptions,markus2025objective}. These assessments often target specific learning interventions and measure the resulting gains within particular activities or curricula. For example, the AI Literacy Concept Inventory (AI-CI) was validated with 981 middle school students and used as a pre-/post-test for participants in the \textit{DAILy} curriculum \cite{zhang2025developing}. Similarly, \citet{iqbal2025teaching} designed two middle school AI awareness modules and evaluated them with pre-/post-test knowledge measures focused on AI misrepresentation. Although assessing specific learning gains is valuable, only a few studies (e.g., \citet{zhang2025developing}) have comprehensively evaluated instrument validity and reliability. The lack of such efforts limits scalability and broader adoption, highlighting the need for rigorous, theory-driven assessment design paired with multifaceted validation (e.g., content, construct, and reliability evidence). One example of such evaluation is the Generative AI Literacy Test (GLAT), which measures factual knowledge and applied skills across four dimensions---Know \& Understand, Use \& Apply, Evaluate \& Create, and Ethics---based on \citeauthor{ng2021conceptualizing}'s \cite{ng2021conceptualizing} AI literacy framework \cite{jin2025glat}. Similarly, \citet{markus2025objective} developed AICOS, an objective AI literacy test synthesized from 15 competency measures, psychometrically tested with a sample of 514 participants, and designed for use in both educational and professional contexts.


Although many AI literacy assessments exist, two key gaps remain: (1) most assessments target students, with few designed specifically for educators \cite{chai2021perceptions}, and (2) the relationship between teachers' self-reported perceptions and their objective performance remains underexplored. To address these gaps, this study develops educator-focused self-reported and objective assessments within a shared framework and examines their validity and interrelationships.

\section{Methods} 

\subsection{Survey Design}  
The survey included 10 demographic items along with responses to 15 self-report items and 25 objective-based items, all aligned with established AI literacy frameworks \cite{ng2021conceptualizing, jin2025glat}.

\subsubsection{Objective-Based Items}
Twenty-five objective-based items were designed or selected to align with \citeauthor{ng2021conceptualizing}'s \cite{ng2021conceptualizing} four dimensions of AI literacy: (1) Know \& Understand AI, (2) Use \& Apply AI, (3) Evaluate \& Create AI, and (4) AI Ethics. These consisted of 25 multiple-choice questions (MCQs) distributed across the four dimensions. For the Know \& Understand AI dimension, two items were adapted from the GLAT AI literacy assessment \cite{jin2025glat}, while the remaining 23 items were iteratively developed by two researchers with expertise in assessment design, following established question design principles and checklists \cite{rush2016impact}. All items were written in a scenario-based format that situates test-takers in realistic pedagogical contexts and requires them to engage with AI to solve educational problems. This approach was adopted because scenario-based assessment is well-suited to measuring the application of knowledge and strategies in authentic contexts, with empirical support for its validity \cite{sabatini2020engineering,darzhinova2025technology}. For example, an item under the Use \& Apply AI dimension assesses the ability to select an appropriate temperature setting when using a generative AI tool, with the correct option bolded:

\begin{quote}\small
\textbf{\LabelText{[Use \& Apply AI]}} Below are classroom tasks related to the game \textit{Identity V}. Please determine which one is the most appropriate to be set as \textbf{high} temperature for the chatbot.

\begin{enumerate}[label=\Alph*., leftmargin=*, itemsep=2pt]
  \item \textbf{Brainstorming new skin designs for your favorite hunter or survivor and asking the chatbot to generate creative and unique ideas.}
  \item Creating a detailed guide about the maps in Identity V by asking for key areas and strategies for each map.
  \item Organizing a list of all hunters' and survivors' abilities to better understand their strengths and weaknesses.
  \item Writing a step-by-step strategy for winning as a hunter in a ranked match.  

\end{enumerate}
\end{quote}

\subsubsection{Self-Report Items} 
Fifteen self-reported items were used to measure teachers' perceived AI literacy competencies on a 5-point Likert scale (1 = strongly disagree; 5 = strongly agree). The items were adapted from AILS-CCS \cite{ma2024development}, a comprehensive scale grounded in a well-established AI literacy framework \cite{ng2021conceptualizing}, with minor modifications to better align with teachers' everyday practice. Example items are shown below:

\begin{quote}\small
\begin{enumerate}[label=, leftmargin=*, itemsep=2pt]
  \item \textbf{\LabelText{[Know \& Understand AI]}} I can distinguish between AI and non-AI devices.
  \item \textbf{\LabelText{[Use \& Apply AI]}} I am skilled in using AI applications to help me complete daily tasks.
  \item \textbf{\LabelText{[Evaluate \& Create AI]}} I can identify when it is beneficial for my students to use AI in their learning.
  \item \textbf{\LabelText{[AI Ethics]}} I am always cautious about the misuse of AI technology. 
\end{enumerate}
\end{quote}

After item generation, a native Traditional Chinese speaker translated all items into Traditional Chinese. A Taiwanese K--12 teacher then piloted the full instrument and provided feedback on wording and cultural nuances in the scenarios. This feedback was incorporated through iterative revisions prior to launching the survey.

\subsubsection{Demographics Items}
We collected demographic and contextual information, including participants' age, gender, grade levels taught, highest degree earned, employment status (in-service/pre-service teacher), years of teaching experience, IT devices provided by the school, whether the school provides IT/CS courses, and whether they had prior AI literacy learning experience.

\subsection{Data Collection and Participants} 
The study received approval from the institutional ethics boards of both the researchers' home institution and local collaborators prior to data collection. Data collection was conducted in February 2025, when the survey was distributed via mailing lists to over 2,000 Taiwanese pre- and in-service K-12 teachers. All participants provided informed consent before completing the survey. Participation was voluntary, and respondents completed the survey through a Google Form. Each participant received a 200 New Taiwan Dollar digital gift card (approximately \$6.6 USD) as compensation.  

A total of 358 participants completed the survey, ranging in age from 18 to 73, with a median age of 33. The sample was 59.4\% female and 40.6\% male. Of the participants, 72.9\% were in-service K--12 teachers and 27.1\% were pre-service teachers. In-service teachers reported an average of 5.3 years of teaching experience. By school level, 49.5\% taught in elementary schools, 17.1\% in middle schools, 18.6\% in high schools, and the remainder served in other roles such as special education. In terms of subject area, nearly half were language teachers, while the others taught mathematics, life sciences, information technology, or social sciences. The highest degree attained was almost evenly split between master’s and bachelor’s degrees. Finally, out of 358 responses, 33.3\% of in-service teachers and 59\% of pre-service teachers reported prior AI literacy learning experiences \footnote{The demographic survey is available on OSF: \newline
\url{https://osf.io/94x2t/?view_only=68270bda63c44cd4b18ed98a49a9c403}}.

\subsection{Data Processing and Analysis} 
Among the total 358 completed surveys, all teachers completed the objective-based items, while 288 completed the self-reported items. After excluding 70 teachers who did not complete the self-report portion, the final analytic sample size is 288. Of the 288 participants, 40.4\% identified as male (the remainder female), 46.5\% were pre-service teachers, and 186 (64.6\%) reported prior AI literacy learning experience.

\subsubsection{Examining Psychometric Stability} 

To examine the psychometric stability of the objective-based AI literacy assessment (\textit{N} = 288), we used a one-parameter logistic Rasch (1PL) model. Following \citet{chiu2024developing}, items were scored dichotomously (1 = correct, 0 = incorrect). The Rasch model was selected over more complex IRT models (e.g., 2PL, 3PL) because the primary aim was to evaluate measurement quality and provide evidence for examining construct validity rather than estimate item discrimination. The model offers interpretable estimates of item difficulty and person ability on a common logit scale, and is widely used in educational measurement for instrument validation \cite{chiu2024developing}. 

Several steps were conducted to evaluate measurement quality. First, internal consistency was estimated using the Kuder–Richardson Formula 20 (KR-20), which is appropriate for measuring the reliability of binary responses, given the dichotomous scoring system of the test \cite{kuder1937theory}. Second, we examined item–person targeting through the Wright Map to determine whether item difficulty aligned with teacher ability levels \cite{boone2016rasch}. Third, item fit statistics (Infit and Outfit mean square values) were used to evaluate the degree to which each item contributed to the underlying construct, where values between 0.5 and 1.5 have been considered acceptable in prior measurement-focused research \cite{baker2001basics, ames2015ncme}. Items that were out of this range were removed. Finally, principal component analysis (PCA) of Rasch residuals was conducted to assess dimensionality, with eigenvalues below 2 interpreted as evidence of approximate unidimensionality \cite{embretson2013item}.

\subsubsection{Factor Analyses} 

We examined the latent structure of the objective-based measures in two stages. First, we conducted an exploratory factor analysis (EFA) to identify the underlying factor structure \cite{topal2025artificial}. Prior to analysis, item-total correlations were calculated, and items with corrected correlations below .30 were removed. Cronbach's $\alpha$ for the full item set was then calculated to assess internal consistency. The suitability of the data for EFA was evaluated using the Kaiser-Meyer-Olkin (KMO) measure of sampling adequacy and Bartlett's test of sphericity. Factors were extracted using common factor analysis with oblimin rotation to allow correlated factors. Factor retention was guided by eigenvalues greater than one, scree plots, and theoretical interpretability. Items with factor loadings below .30 were removed, and Cronbach's $\alpha$ for each factor was calculated at each round of refinement.  

In the second stage, we then conducted a confirmatory factor analysis (CFA) to test the hypothesized factor structures derived from EFA. CFA models were estimated in the full sample (\textit{N} = 288). Model evaluation focused on global fit indices, including the chi-square test of model fit ($\chi^2$), the chi-square to degrees of freedom ratio ($\chi^2/df$), the Comparative Fit Index (CFI), the Tucker-Lewis Index (TLI), the Standardized Root Mean Square Residual (SRMR), and the Root Mean Square Error of Approximation (RMSEA). Following the guidelines from previous literature, $\chi^2/df$ values less than 3 were considered indicative of acceptable fit \cite{brown2015confirmatory}. Values of CFI and TLI above .90 were interpreted as acceptable, and above .95 as good \cite{brown2015confirmatory}. RMSEA and SRMR values below .08 were considered acceptable, with values below .05 indicating close fit \cite{maydeu2013goodness}. Standardized factor loadings greater than .30 were considered acceptable, and internal consistency reliability was evaluated using Cronbach's $\alpha$ with a threshold of .60 \cite{hair2011pls}.  

For the self-reported measures, one negatively worded item was frist reverse-coded, and we then directly conducted a CFA. Since the items were adapted from an established framework in the literature \cite{ma2024development}, we hypothesized that the number of factors would remain consistent with the theoretical model, so we did not test it with an EFA.

\vspace{-2mm}

\subsubsection{Latent Profile Analysis} 

To identify latent learner profiles, we conducted latent profile analysis (LPA) using both the self-reported and objective-based factors identified in the preceding CFA. Prior to analysis, all indicators were standardized, and Gaussian mixture models were applied to extract increasing numbers of candidate profiles (from two-to-eight) for which goodness-of-fit metrics were used for final profile selection. Model selection was guided by the Akaike Information Criterion (AIC) and the Bayesian Information Criterion (BIC), with lower values indicating a better fit. The optimal number of profiles was determined by comparing AIC and BIC values across candidate solutions. Posterior probabilities from the selected models were then used to assign participants to their most likely profile. To evaluate classification quality, we calculated each participant's maximum posterior probability (confidence) and entropy (uncertainty) and summarized these metrics within profiles. We also examined mean scores for each factor across profiles and plotted the standardized profile means with 95\% confidence intervals to support interpretation. This procedure was applied to the full sample (\textit{N} = 288) as well as separately to the subgroups of teachers with prior AI literacy experience (\textit{n} = 186) and those without such experience (\textit{n} = 102) to explore whether prior exposure moderated discrepancies or consistencies. Finally, profiles were qualitatively compared across the three groups by three experts with backgrounds in educational technology, human–AI interaction, and cognitive science. Experts independently reviewed the profiles to identify consistent patterns of underestimation, overestimation, or alignment between self-reports and objective measures. Any discrepancies were discussed.

\section{Results} 

\subsection{To what extent does the objective measure-based AI literacy assessment demonstrate psychometric stability?}

\begin{figure}[t]
  \centering
  \includegraphics[width=0.4\textwidth]{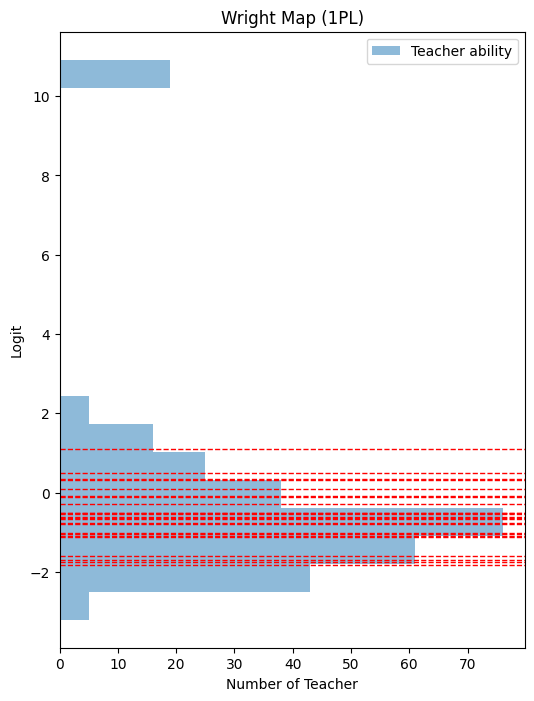}
  \caption{Wright Map of the AI Literacy Objective Measure (1PL Rasch Model).}
  \label{fig:RaschModel}
  \footnotesize \textit{Note.} The blue histogram shows the distribution of teacher ability estimates, and the red dashed lines indicate the estimated difficulty of each test item. 
\end{figure}

The Rasch model analysis indicated that the objective-based AI literacy assessment demonstrated good psychometric properties. Internal consistency was high (KR-20 = 0.862), which indicates reliable measurement. Item-person targeting, as illustrated in the Wright Map (see Figure~\ref{fig:RaschModel}), showed that item difficulties were well aligned with teacher ability levels, with most participants falling between $-2$ and $+2$ logits ($M = -0.60$, $SD = 0.75$). A small number of respondents (\textit{n} = 19) achieved extreme ability estimates above 5 logits, which indicates a potential ceiling effect and suggests that more difficult items may be needed to better differentiate the highest-performing teachers. Item difficulty estimates ranged from $-1.83$ to $1.1$ logits, covering the ability range of most participants. After removing five items that were outside the acceptable range for fit statistics (0.5-1.5), the remaining items demonstrated acceptable Rasch model fit, with Infit statistics ranging from 0.78 to 1.28 and Outfit statistics ranging from 0.63 to 1.45. This indicates that each item contributed meaningfully to the construct being measured. The PCA of Rasch residuals supported approximate unidimensionality, with the eigenvalue of the first contrast equal to 0.44 and all subsequent contrasts below 2.0. Overall, the Rasch results suggest that the objective-based AI literacy assessment is psychometrically stable, with high reliability, appropriate item-person targeting, good item fit, and support for unidimensionality.

\subsection{What factors emerge from K-12 teachers' self-reported and objectively-measured levels of AI literacy?}

\subsubsection{Objective-based measure} After removing 9 items with corrected item–total correlations all below .30 (including 5 items removed from the Rasch model of RQ1 results), the remaining 20 items had correlations ranging from .32 to .61, which indicates adequate internal consistency. 

Sampling adequacy for the objective-based measure was strong, with a Kaiser-Meyer-Olkin (KMO) value of 0.873, which is well above the recommended cutoff of .60 from the literature \cite{kaiser1974index}, indicating that the data were appropriate for factor analysis. Bartlett's test of sphericity was also significant ($\chi^2 = 1322.13$, $p < .001$), suggesting that the correlation matrix was not an identity matrix and thus factorable. The exploratory factor analysis with oblimin rotation initially produced four factors with eigenvalues greater than one (5.39, 1.51, 1.24, and 1.09). However, one factor included only a single item, which would prevent it from being evaluated for reliability, and the scree plot showed a clear leveling after the third factor. Therefore, we finalized the EFA with a three-factor structure. Items with loadings below .30 were removed, and this process resulted in a final structure of 18 items grouped across three distinct but correlated factors. Each factor demonstrated internal consistency reliability above the .60 threshold, indicating that the items within each dimension cohered well and contributed meaningfully to the underlying construct.

\begin{figure}[t]
  \centering
  \includegraphics[width=0.4\textwidth]{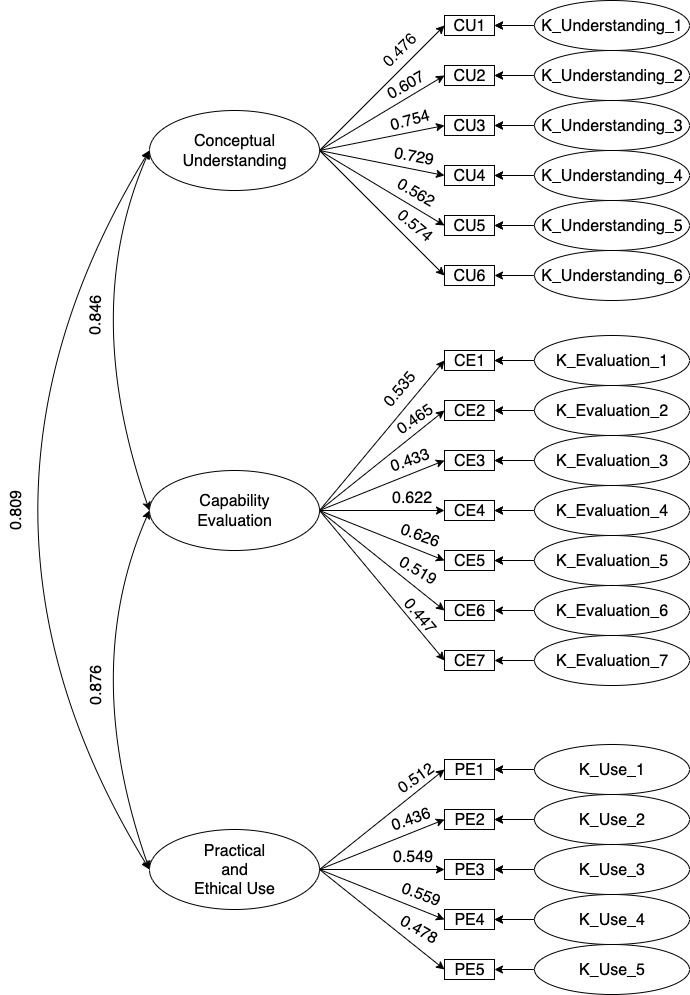}
  \caption{Confirmatory Factor Analysis of the AI Literacy Objective Measure with Three Factors}
  \label{Figure:AI_Literacy_OM}
\end{figure}

Subsequently, CFA was conducted on the three-factor model identified from the EFA in the full sample. As shown in Figure~\ref{Figure:AI_Literacy_OM}, the model demonstrated acceptable fit in the full sample (\textit{N} = 288), $\chi^2(132) = 228.33$, $p < .001$; $\chi^2/df = 1.73$; CFI = .914; TLI = .901; RMSEA = .05, SRMR = 0.055. Standardized factor loadings ranged from .43 to .75, and factor reliabilities were $\alpha = .72$, .78, and .63 for the three factors. The Cronbach's alpha for the 18-item scale was $\alpha = 0.85$. Overall, the factor analyses showed that the objective-based measure was psychometrically stable. A three-factor structure was identified and confirmed, with the model fitting reasonably well.



\subsubsection{Self-reported measure} 
\begin{figure}[t]
  \centering
  \includegraphics[width=0.4\textwidth]{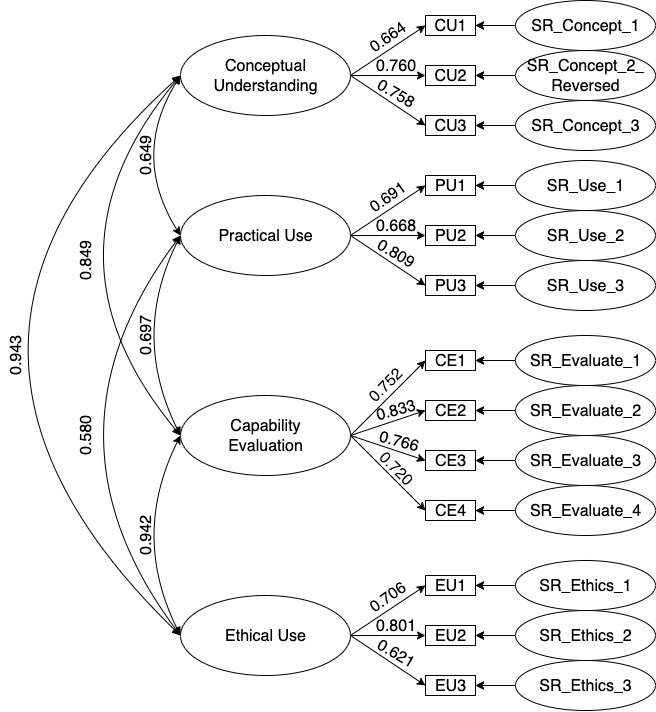}
  \caption{Confirmatory Factor Analysis of the AI Literacy Self-Report Measure with Four Factors}
  \label{Figure:AI_Literacy_SR}
\end{figure}

We first reverse-coded the single negatively worded item and specified a four-factor model informed by the theoretical framework (Concept, Use, Evaluate, and Ethics). The initial CFA with 16 items demonstrated marginal fit ($\chi^2(98) = 278.64$, $p < .001$; $\chi^2/df = 2.84$; CFI = .907; TLI = .886; RMSEA = .080). Standardized factor loadings ranged from .43 to .83, with most items exceeding .60. Internal consistency was acceptable across all four factors: Concept ($\alpha = .77$), Use ($\alpha = .76$), Evaluate ($\alpha = .82$), and Ethics ($\alpha = .71$). To improve fit, three items with standardized loadings below .50 were removed, resulting in a final 13-item model. This optimized model demonstrated good fit in the full sample (\textit{N} = 288; $\chi^2(59) = 109.29$, $p < .001$; $\chi^2/df = 1.85$; CFI = .970; TLI = .961; RMSEA = .054; SRMR = 0.04). Standardized factor loadings ranged from .62 to .83, and internal consistency was adequate across all factors: Concept ($\alpha = .77$), Use ($\alpha = .76$), Evaluate ($\alpha = .85$), and Ethics ($\alpha = .75$), as shown in Figure~\ref{Figure:AI_Literacy_SR}. The Cronbach's alpha for the 13-item overall was $\alpha = 0.889$. These results support internal consistency and convergent validity, with a theoretically coherent four-factor structure sufficiently represented by a refined 13-item model \footnote{Finalized items are available on OSF: \newline
\url{https://osf.io/94x2t/?view_only=68270bda63c44cd4b18ed98a49a9c403}}.

\subsubsection{Weak correlations between objective-based and self-reported measures}  
Correlations between the three OB factors---\textit{Conceptual Understanding\_OB}, \textit{Capability Evaluation\_OB}, and \textit{Practical and Ethical Use\_OB}---and the four SR factors were consistently weak, ranging from $r = 0.07$ to $r = 0.24$. Specifically, \textit{Concept\_SR} correlated with \textit{Conceptual Understanding\_OB} ($r = 0.24$), \textit{Capability Evaluation\_OB} ($r = 0.14$), and \textit{Practical and Ethical Use\_OB} ($r = 0.14$). \textit{Ethics\_SR} correlated with \textit{Conceptual Understanding\_OB} ($r = 0.23$), \textit{Capability Evaluation\_OB} ($r = 0.17$), and \textit{Practical and Ethical Use\_OB} ($r = 0.11$). \textit{Evaluate\_SR} showed weaker correlations with \textit{Conceptual Understanding\_OB} ($r = 0.17$), \textit{Capability Evaluation\_OB} ($r = 0.10$), and \textit{Practical and Ethical Use\_OB} ($r = 0.07$). Finally, \textit{Use\_SR} correlated with \textit{Conceptual Understanding\_OB} ($r = 0.18$), \textit{Capability Evaluation\_OB} ($r = 0.10$), and \textit{Practical and Ethical Use\_OB} ($r = 0.10$). Overall, these uniformly low coefficients suggest that teachers’ self-reported competencies are weakly associated with their knowledge-based performance.

\subsection{What distinct learner profiles emerge from the self-report- and objective measure-based factors?}

\begin{figure}[t]
  \centering
  \includegraphics[width=0.6\textwidth]{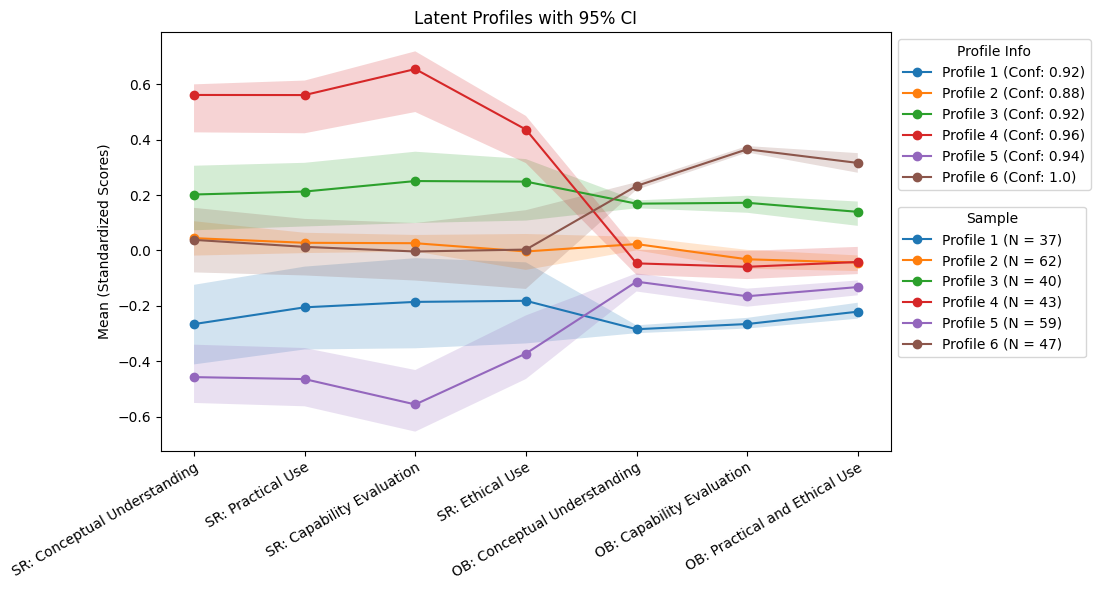}
  \caption{Latent Profile Analysis of Combined Self-Reported and Objective-Based AI Literacy (N=288)}
  \label{Figure:288LPAwith6C}
\end{figure}

LPA was conducted on the combined self-reported and objective-based factors in the full sample (\textit{N} = 288). Model selection using both AIC and BIC indicated that the six-profile solution provided the best fit. The emergent profiles varied meaningfully in terms of both self-reported and objective-based AI literacy. As shown in Figure~\ref{Figure:288LPAwith6C}, Profile~4 (\textit{n} = 43) reflected teachers who rated themselves consistently high across self-reported dimensions but had lower scores on objective measures, which suggested overestimation of competence. In contrast, Profile~3 (\textit{n} = 40) showed alignment between moderately high self-reports and similarly high objective scores, while Profile~5 (\textit{n} = 59) reflected the opposite trend—low self-reports despite near-average objective performance—which indicated underestimation. Profile~1 (\textit{n} = 37) and Profile~2 (\textit{n} = 62) clustered near the overall mean on both self-reported and objective measures, with little discrepancy between perception and performance. Finally, Profile~6 (\textit{n} = 47) demonstrated relatively strong objective-based competencies with average self-reports, which suggested balanced but more accurate knowledge.

Classification quality was high, with an overall mean confidence of .935 and a low mean entropy of .25. Profile-specific confidence values ranged from .88 (Profile~2) to 1.00 (Profile~6), with correspondingly entropy values (.013–.466), indicating reliable assignment of participants into distinct profiles. The six profiles reveal systematic differences between teachers' perceived and demonstrated AI literacy, ranging from strong alignment to notable under- and overestimation.

\subsection{How do these profiles differ among those with prior AI literacy experience and those without?}

\begin{figure}[t]
  \centering
  \begin{subfigure}[t]{0.6\textwidth}
    \centering
    \includegraphics[width=\textwidth]{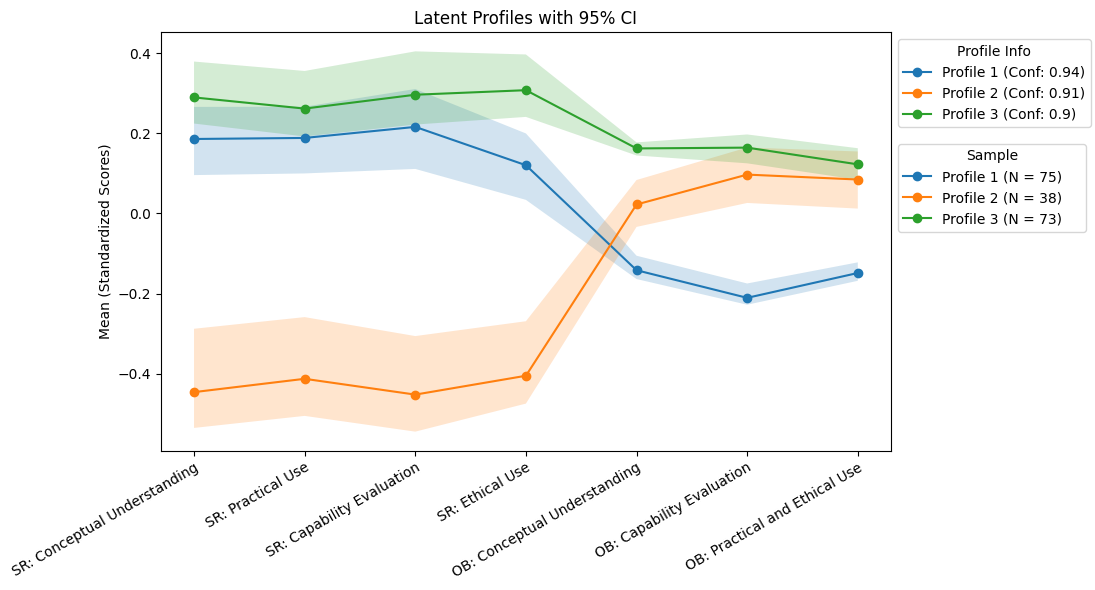}
    \caption{Teachers with Prior AI Literacy Experience (N=186)}
    \label{Figure:186LPAwith3C}
  \end{subfigure}\hfill
  \begin{subfigure}[t]{0.6\textwidth}
    \centering
    \includegraphics[width=\textwidth]{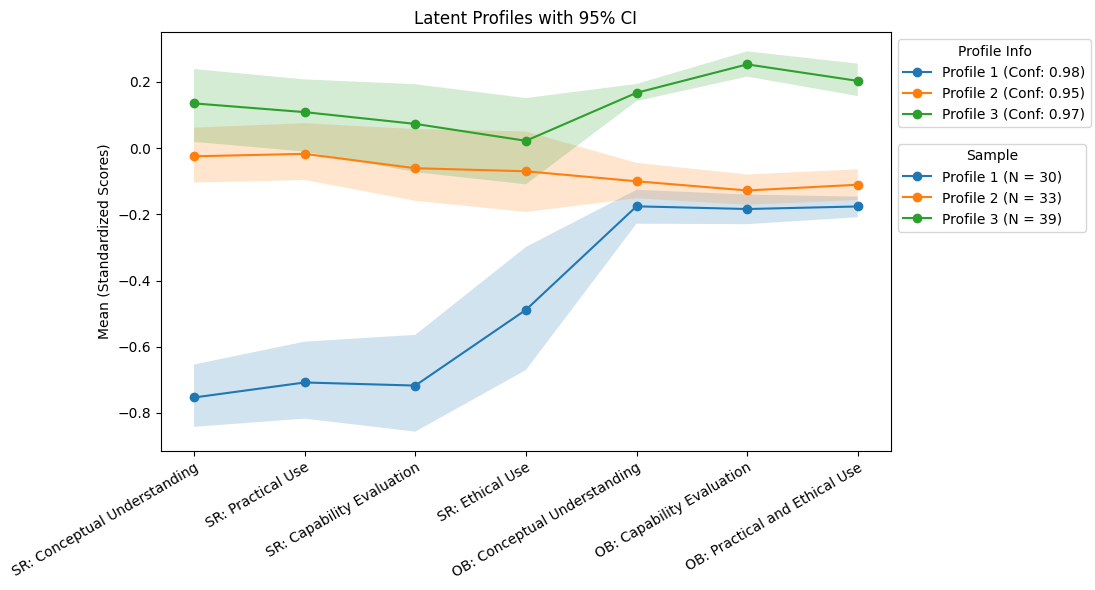}
    \caption{Teachers without Prior AI Literacy Experience (N=102)}
    \label{Figure:102LPAwith3C}
  \end{subfigure}
  \caption{Latent Profile Analyses of combined self-reported and objective-based AI literacy by prior AI-literacy experience.}
  \label{fig:LPA-comparison}
\end{figure}

For teachers with prior AI literacy experience (\textit{N} = 186), the analysis supported a three-profile solution. As shown in Figure~\ref{Figure:186LPAwith3C}, Profile~1 (\textit{n} = 75) included teachers who rated themselves slightly above average on the self-report factors but scored closer to average or below on the objective measures, which reflected mild overestimation between perception and performance. Profile~2 (\textit{n} = 38) comprised teachers with consistently low self-ratings across all four dimensions while achieving average performance on the objective measures, and this pattern reflects underestimation. Profile~3 (\textit{n} = 73) represented teachers with higher self-ratings that matched moderately strong objective scores, and this alignment reflects more accurate self-perceptions. Moreover, profile-specific classification quality was high, with confidence values of .936 (Profile 0), .909 (Profile 1), and .905 (Profile 2). The corresponding entropy values were .232, .345, and .336. Overall, the average classification confidence reached .918, and the mean entropy was .296.


For teachers without prior AI literacy experience (\textit{N} = 102), the analysis indicated a three-profile solution. As shown in Figure~\ref{Figure:102LPAwith3C}, Profile~1 (\textit{n} = 30) included teachers with low self-reported ratings across all four dimensions and weak performance below the average on the objective measures. Profile~2 (\textit{n} = 33) represented teachers whose self-ratings were close to the average but whose objective scores fell slightly below average, showing only a modest gap between perception and performance. Profile~3 (\textit{n} = 39) captured teachers with higher self-ratings that corresponded with strong objective scores across all factors, and this reflects close alignment between confidence and competence. Classification quality was high, with an overall confidence of .963 and entropy of .131, and profile-specific confidence values of .98, .95, and .97 that confirmed clear separation among the three groups.


When comparing the two subgroups, overestimation (high self-reports paired with lower performance) appeared only among teachers with prior AI literacy experience, whereas underestimation emerged in both groups. Specifically, in the experienced group, Profile~1 (\textit{n} = 75) reflected mild overestimation, with self-reports slightly above average but objective scores closer to average or below, while Profile~2 (\textit{n} = 38) reflected underestimation, with consistently low self-reports despite average objective performance. In contrast, among teachers without prior experience, Profile~1 (\textit{n} = 30) showed a low–low pattern, with both self-reports and objective scores well below average, alongside evidence of underestimation. This low–low profile did not appear in the experienced group.


\vspace{-1mm}

\section{Discussion} 
This study establishes psychometrically stable instruments for assessing K–12 teachers’ AI literacy using both self-report and objective-based measures. Using responses from 288 teachers, we examined instrument quality, factor structures, and learner profiles. These analyses yield key insights that together provide a new perspective on teachers’ AI literacy assessment.

\subsection{Psychometric Validation of OB and SR Measures of AI Literacy}

The Rasch analysis of the OB assessment demonstrated high reliability (KR-20 = .862), appropriate item–person targeting, acceptable item fit after the removal of five misfitting items, and approximate unidimensionality. Factor analyses further clarified the structure. For the OB assessment, three factors emerged with 18 items, \textit{Conceptual Understanding}, \textit{Capability Evaluation}, and \textit{Practical/Ethical Use}, all with acceptable reliabilities and standardized loadings above .43. For the SR scale, CFA supported a refined 13-item, four-factor model capturing teachers' self-perceptions of \textit{Concept}, \textit{Use}, \textit{Evaluate}, and \textit{Ethics}. This model showed strong reliability ($\alpha = .77$--$.85$) and good global fit indices. By validating both instruments, this study extends prior work that has typically examined either objective measures \cite{chiu2024developing, markus2025objective} or self-reported measures\cite{ma2024development, carolus2023mails}. It demonstrates both teachers' objectively measured AI literacy competencies and their self-reported perceptions, using a shared framework across \textit{Understanding}, \textit{Evaluation}, \textit{Use and Apply}, and \textit{Ethics}, thereby providing a more comprehensive and integrated picture of teachers' AI literacy.

\vspace{-1.5mm}
\subsection{Profiles of Alignment and Misalignment Between SR and OB}
Beyond measurement, LPA illustrates how SR and OB factors combine to reveal systematic patterns of alignment and misalignment between teachers' perception and demonstrated competence, and how prior AI literacy exposure shapes these patterns. In the full sample, six distinct profiles appeared, spanning overestimation (high SR factors with weaker OB performance), underestimation (low SR factors with average OB performance), and alignment (SR and OB factors at comparable levels). Subgroup analyses showed clearer contrasts. Among teachers with prior AI literacy experience (\textit{n} = 186), Profile~1 reflected mild overestimation, with SR scores on Concept, Use, Evaluate, and Ethics slightly above average but OB scores on Conceptual Understanding, Capability Evaluation, and Practical/Ethical Use near or below average. Profile~2 captured underestimation, with consistently low SR scores but average or above OB performance. In contrast, among teachers without prior experience (\textit{n} = 102), Profile~1 represented a unique low–low pattern, with both SR and OB scores well below average. This comparison highlights two key differences: (1) the low–low pattern was unique to teachers without prior AI literacy experience, and (2) overestimation among experienced teachers appeared milder and more calibrated than in the inexperienced group. These results demonstrate how SR and OB measures can be used together to diagnose calibration issues and show that prior AI literacy education is associated with fewer extreme mismatches and more balanced self-assessment.  

These results connect directly to core themes in LA. First, they demonstrate how SR and OB measures can be combined to highlight gaps between learners’ perceived and demonstrated competence. Recognizing these gaps is important for understanding where learners may misjudge their abilities and for informing the design of interventions on such discrepancies. Second, the profile-based approach illustrates how model-based clustering (via LPA) can uncover \textit{heterogeneity in teacher learning trajectories}, supporting the LA goal of tailoring interventions to distinct learner subgroups. Third, by validating instruments that can be embedded into teacher professional development programs, this study shows how LA can move beyond post-hoc evaluation to become part of an \textit{adaptive feedback loop}: diagnostic assessments inform targeted supports, and subsequent data collection tracks growth over time. Finally, the explicit incorporation of the \textit{Ethics} dimension in both SR and OB instruments aligns with broader LA discussions on responsible AI and the need to foreground equity, fairness, and societal impacts when analyzing and acting on learner data.

\subsection{Extending AI Literacy Frameworks with SR–OB Validation for Evidence-Based PD}
Our findings carry both theoretical and practical implications. This study advances the measurement of teacher AI literacy by building on and extending prior frameworks. For example, \citet{mills2024ai} proposed a self-report framework centered on conceptual, technical, and pedagogical dimensions of AI literacy for educators, while \citet{ng2021conceptualizing} synthesized 30 studies into four broad dimensions of knowing/understanding, using/applying, evaluating/creating, and ethics. Similarly, \citet{chiu2024developing} emphasized the need to move beyond self-reported perceptions toward validated objective measures for K–12 learners. Our study expands on these works in two important ways. First, we validated both self-report and objective-based instruments within a shared framework, enabling a systematic comparison between teachers' perceptions and demonstrated competencies, contributing to future work of building richer learner models. Second, we explicitly incorporated \textit{Ethics} as a dimension in both SR and OB measures, extending Mills' educator-focused framework and aligning with \citeauthor{ng2021conceptualizing}'s \cite{ng2021conceptualizing} and \citeauthor{chiu2024developing}'s \cite{chiu2024developing} emphasis on the social and ethical implications of AI. Moreover, our findings show that OB and SR are not correlated and cannot be used interchangeably to represent teachers’ AI literacy. This echoes recent findings showing discrepancies between self-assessment scales and performance-based tests on AI Literacy, which may reflect metacognitive biases such as the Dunning–Kruger effect \cite{bewersdorff2025ai}. Practically, the validated SR and OB measures serve as diagnostic tools that can be embedded into AI literacy PD programs, both before and after training, to assess changes in teachers' perceived perceptions and demonstrated competencies by monitoring teachers' growth, and detecting calibration issues. Insights from these assessments can further inform the design of differentiated professional development or targeted scaffolding for specific subgroups of teachers to ensure that support is responsive to patterns of overestimation, underestimation, or alignment. By bridging psychometric rigor with LA methods, this study contributes to advancing the design of AI literacy interventions that are both evidence-based and responsive to the diverse needs of educators.

\section{Limitations and Future Work} 
This study has several limitations that point to directions for future research. First, the analysis did not separate teachers by their pre-service or in-service status, and their teaching experience, which may limit the ability to examine how profiles may differ across these demographic factors and individual differences. Future work could consider including this information to provide a more nuanced understanding of AI literacy patterns. Second, although the objective-based assessment used in this study was scenario-based, it was not tailored to specific subjects or grade levels. As a next step, we plan to design subject-specific and grade-divided (primary and secondary) scenario-based items that can be embedded into PD programs. Such tools would allow for automatic identification of teachers' strengths and weaknesses and provide differentiated scaffolding for subgroups of teachers. Third, recent AI literacy frameworks have begun to include dimensions such as Detect AI \cite{carolus2023mails} and Generative AI literacy \cite{markus2025objective}. While we acknowledge the value of these developments, many existing items are overly technical for K–12 educators. For example, in \citeauthor{jin2025glat}'s \cite{jin2025glat} Generative AI Literacy Test, some items focus on retrieval-augmented generation and tokenization. Although such items can differentiate advanced ability, most K–12 teachers—without systematic training—lack the background to understand or accurately answer these questions. As a result, these measures may capture unfamiliarity with technical terminology rather than the capacity to integrate AI into practice. In the meantime, teachers do not need to know everything about AI; rather, they need sufficient knowledge and skills to confidently integrate AI into classrooms. No single PD can provide everything, and the specific knowledge required will depend on the type of PD teachers pursue.


\vspace{-2mm}
\section{Conclusion} 
This study contributes to the growing field of AI literacy in education by evaluating the validity of both self-report and objective-based measures through psychometric testing. The measures share a set of dimensions using an established framework and reveal distinct profiles that highlight patterns of alignment and misalignment between teachers' perceptions and demonstrated competencies. Results revealed the importance of considering both types of measures together to capture a more complete picture of teachers' AI literacy. Our findings also suggest that prior AI literacy experience plays a role in reducing extreme mismatches and fostering more balanced self-assessment. In addition, this work advances LA by demonstrating how validated instruments and profile-based analyses can yield interpretable insights for monitoring growth, detecting calibration gaps, and supporting adaptive feedback loops. By uncovering heterogeneity in teachers’ learning trajectories, our study contributes to the goals of enhancing personalization and scalability, as well as embedding diagnostic assessments into professional development and LA-focused ecosystems.



\begin{acks}
We would like to thank the National Science Foundation (\#2331379), the Gates Foundation, and other anonymous philanthropies.
\end{acks}

\bibliographystyle{ACM-Reference-Format}
\bibliography{samples/reference}

@article{ng2021conceptualizing,
  title={Conceptualizing AI literacy: An exploratory review},
  author={Ng, Davy Tsz Kit and Leung, Jac Ka Lok and Chu, Samuel Kai Wah and Qiao, Maggie Shen},
  journal={Computers and Education: Artificial Intelligence},
  volume={2},
  pages={100041},
  year={2021},
  publisher={Elsevier}
}

@article{ng2023teachers,
  title={Teachers’ AI digital competencies and twenty-first century skills in the post-pandemic world},
  author={Ng, Davy Tsz Kit and Leung, Jac Ka Lok and Su, Jiahong and Ng, Ross Chi Wui and Chu, Samuel Kai Wah},
  journal={Educational technology research and development},
  volume={71},
  number={1},
  pages={137--161},
  year={2023},
  publisher={Springer}
}

@article{ning2024teachers,
  title={Teachers’ AI-TPACK: Exploring the relationship between knowledge elements},
  author={Ning, Yimin and Zhang, Cheng and Xu, Binyan and Zhou, Ying and Wijaya, Tommy Tanu},
  journal={Sustainability},
  volume={16},
  number={3},
  pages={978},
  year={2024},
  publisher={MDPI}
}

@inproceedings{polak2022teachers,
  title={Teachers’ perspective on artificial intelligence education: An initial investigation},
  author={Polak, Sara and Schiavo, Gianluca and Zancanaro, Massimo},
  booktitle={CHI conference on human factors in computing systems extended abstracts},
  pages={1--7},
  year={2022}
}

@article{yue2024understanding,
  title={Understanding K--12 teachers’ technological pedagogical content knowledge readiness and attitudes toward artificial intelligence education},
  author={Yue, Miao and Jong, Morris Siu-Yung and Ng, Davy Tsz Kit},
  journal={Education and information technologies},
  volume={29},
  number={15},
  pages={19505--19536},
  year={2024},
  publisher={Springer}
}

@article{keefer2015self,
  title={Self-report assessments of emotional competencies: A critical look at methods and meanings},
  author={Keefer, Kateryna V},
  journal={Journal of Psychoeducational Assessment},
  volume={33},
  number={1},
  pages={3--23},
  year={2015},
  publisher={Sage Publications Sage CA: Los Angeles, CA}
}

@article{xiao2025improving,
  title={Improving Student-AI Interaction Through Pedagogical Prompting: An Example in Computer Science Education},
  author={Xiao, Ruiwei and Hou, Xinying and Ye, Runlong and Kazemitabaar, Majeed and Diana, Nicholas and Liut, Michael and Stamper, John},
  journal={arXiv preprint arXiv:2506.19107},
  year={2025}
}

@article{chiu2024developing,
  title={Developing and validating measures for AI literacy tests: From self-reported to objective measures},
  author={Chiu, Thomas KF and Chen, Yifan and Yau, King Woon and Chai, Ching-sing and Meng, Helen and King, Irwin and Wong, Savio and Yam, Yeung},
  journal={Computers and Education: Artificial Intelligence},
  volume={7},
  pages={100282},
  year={2024},
  publisher={Elsevier}
}

@article{jin2025glat,
  title={GLAT: The generative AI literacy assessment test},
  author={Jin, Yueqiao and Martinez-Maldonado, Roberto and Ga{\v{s}}evi{\'c}, Dragan and Yan, Lixiang},
  journal={Computers and Education: Artificial Intelligence},
  pages={100436},
  year={2025},
  publisher={Elsevier}
}

@article{wang2023k,
  title={K-12 Education in the Age of AI: A Call to Action for K-12 AI Literacy},
  author={Wang, Ning and Lester, James},
  journal={International journal of artificial intelligence in education},
  volume={33},
  number={2},
  pages={228--232},
  year={2023},
  publisher={Springer}
}

@article{chai2021perceptions,
  title={Perceptions of and behavioral intentions towards learning artificial intelligence in primary school students},
  author={Chai, Ching Sing and Lin, Pei-Yi and Jong, Morris Siu-Yung and Dai, Yun and Chiu, Thomas KF and Qin, Jianjun},
  journal={Educational Technology \& Society},
  volume={24},
  number={3},
  pages={89--101},
  year={2021},
  publisher={JSTOR}
}

@article{nazaretsky2022teachers,
  title={Teachers' trust in AI-powered educational technology and a professional development program to improve it},
  author={Nazaretsky, Tanya and Ariely, Moriah and Cukurova, Mutlu and Alexandron, Giora},
  journal={British journal of educational technology},
  volume={53},
  number={4},
  pages={914--931},
  year={2022},
  publisher={Wiley Online Library}
}

@article{mikeladze2024comprehensive,
  title={A comprehensive exploration of artificial intelligence competence frameworks for educators: A critical review},
  author={Mikeladze, Tamar and Meijer, Paulien C and Verhoeff, Roald P},
  journal={European Journal of Education},
  volume={59},
  number={3},
  pages={e12663},
  year={2024},
  publisher={Wiley Online Library}
}

@book{ng2022ai,
  title={AI literacy in K-16 classrooms},
  author={Ng, Davy Tsz Kit and Leung, Jac Ka Lok and Su, Maggie Jiahong and Yim, Iris Heung Yue and Qiao, Maggie Shen and Chu, Samuel Kai Wah},
  year={2022},
  publisher={Springer}
}

@article{ning2025development,
  title={Development and validation of the artificial intelligence literacy scale for teachers (AILST)},
  author={Ning, Yimin and Zhang, Wenjun and Yao, Dengming and Fang, Bowen and Xu, Binyan and Wijaya, Tommy Tanu},
  journal={Education and Information Technologies},
  pages={1--35},
  year={2025},
  publisher={Springer}
}

@article{tenberga2024artificial,
  title={Artificial intelligence literacy competencies for teachers through self-assessment tools},
  author={Tenberga, Ieva and Daniela, Linda},
  journal={Sustainability},
  volume={16},
  number={23},
  pages={10386},
  year={2024},
  publisher={MDPI}
}

@article{younis2025artificial,
  title={The artificial intelligence literacy (AIL) scale for teachers: A tool for enhancing AI education},
  author={Younis, Bilal},
  journal={Journal of Digital Learning in Teacher Education},
  volume={41},
  number={1},
  pages={37--56},
  year={2025},
  publisher={Taylor \& Francis}
}

@inproceedings{hutchins2025empowering,
  title={Empowering Educators in AI: Insights from Co-Designing an AI Microcredential with and for K-12 Educators},
  author={Hutchins, Nicole M and Zhang, Shan and Barrett, Joanne R and Isreal, Maya},
  booktitle={Proceedings of the AAAI Conference on Artificial Intelligence},
  volume={39},
  number={28},
  pages={29137--29144},
  year={2025}
}

@online{TeachAI_Toolkit,
  author    = {{TeachAI}},
  title     = {AI Guidance for Schools Toolkit},
  year      = {2023},
  url       = {https://www.teachai.org/toolkit},
  urldate   = {2025-09-16}
}

@online{i2learning,
  author  = {{i2 Learning}},
  title   = {i2 Learning},
  year    = {2025},
  url     = {https://i2learning.org/},
  urldate = {2025-09-16}
}

@inproceedings{breazeal2023day,
  title={Day of AI: Innovating Pedagogical Practices to Bring AI Literacy to Classrooms at Scale},
  author={Breazeal, Cynthia and Du, Xiaoxue and Abelson, Hal and Klopfer, Eric and Park, Hae Won},
  booktitle={International Conference on Artificial Intelligence in Education Technology},
  pages={267--281},
  year={2023},
  organization={Springer}
}

@misc{codeorgTeachers2025,
  author       = {{Code.org}},
  title        = {Teachers – Free K–12 Computer Science \& AI Curriculum and Training},
  year         = {2025},
  howpublished = {Code.org website},
  url          = {https://code.org/en-US/teachers}
}

@article{zhang2025learning,
  title={Learning About AI: A Systematic Review of Reviews on AI Literacy},
  author={Zhang, Shan and Ganapathy Prasad, Priyadharshini and Schroeder, Noah L},
  journal={Journal of Educational Computing Research},
  pages={07356331251342081},
  year={2025},
  publisher={SAGE Publications Sage CA: Los Angeles, CA}
}

@article{song2024framework,
  title={A framework for inclusive AI learning design for diverse learners},
  author={Song, Yukyeong and Weisberg, Lauren R and Zhang, Shan and Tian, Xiaoyi and Boyer, Kristy Elizabeth and Israel, Maya},
  journal={Computers and Education: Artificial Intelligence},
  volume={6},
  pages={100212},
  year={2024},
  publisher={Elsevier}
}

@inproceedings{vyortkina2024ai,
  title={AI Literacy Framework for Educators: Challenges and Opportunities},
  author={Vyortkina, Dina},
  booktitle={Society for Information Technology \& Teacher Education International Conference},
  pages={898--903},
  year={2024},
  organization={Association for the Advancement of Computing in Education (AACE)}
}

@article{ma2024development,
  title={The development and validation of the artificial intelligence literacy scale for Chinese college students (AILS-CCS)},
  author={Ma, Shuai and Chen, Zhenzhen},
  journal={Ieee Access},
  year={2024},
  publisher={IEEE}
}

@article{markus2025objective,
  title={Objective Measurement of AI Literacy: Development and Validation of the AI Competency Objective Scale (AICOS)},
  author={Markus, Andr{\'e} and Carolus, Astrid and Wienrich, Carolin},
  journal={arXiv preprint arXiv:2503.12921},
  year={2025}
}

@article{carolus2023mails,
  title={MAILS-Meta AI literacy scale: Development and testing of an AI literacy questionnaire based on well-founded competency models and psychological change-and meta-competencies},
  author={Carolus, Astrid and Koch, Martin J and Straka, Samantha and Latoschik, Marc Erich and Wienrich, Carolin},
  journal={Computers in Human Behavior: Artificial Humans},
  volume={1},
  number={2},
  pages={100014},
  year={2023},
  publisher={Elsevier}
}

@article{hair2011pls,
  title={PLS-SEM: Indeed a silver bullet},
  author={Hair, Joe F and Ringle, Christian M and Sarstedt, Marko},
  journal={Journal of Marketing theory and Practice},
  volume={19},
  number={2},
  pages={139--152},
  year={2011},
  publisher={Taylor \& Francis}
}

@book{baker2001basics,
  title={The basics of item response theory},
  author={Baker, Frank B},
  year={2001},
  publisher={ERIC}
}

@book{brown2015confirmatory,
  title={Confirmatory factor analysis for applied research},
  author={Brown, Timothy A},
  year={2015},
  publisher={Guilford publications}
}

@article{ames2015ncme,
  title={An NCME instructional module on item-fit statistics for Item Response Theory models},
  author={Ames, Allison J and Penfield, Randall D},
  journal={Educational Measurement: Issues and Practice},
  volume={34},
  number={3},
  pages={39--48},
  year={2015},
  publisher={Wiley Online Library}
}

@article{maydeu2013goodness,
  title={Goodness-of-fit assessment of item response theory models},
  author={Maydeu-Olivares, Alberto},
  journal={Measurement: Interdisciplinary Research and Perspectives},
  volume={11},
  number={3},
  pages={71--101},
  year={2013},
  publisher={Taylor \& Francis}
}

@book{embretson2013item,
  title={Item response theory for psychologists},
  author={Embretson, Susan E and Reise, Steven P},
  year={2013},
  publisher={Psychology Press}
}

@article{mills2024ai,
  title={AI Literacy: A Framework to Understand, Evaluate, and Use Emerging Technology.},
  author={Mills, Kelly and Ruiz, Pati and Lee, Keun-woo and Coenraad, Merijke and Fusco, Judi and Roschelle, Jeremy and Weisgrau, Josh},
  journal={Digital Promise},
  year={2024},
  publisher={ERIC}
}

@article{chiu2021holistic,
  title={A holistic approach to the design of artificial intelligence (AI) education for K-12 schools},
  author={Chiu, Thomas KF},
  journal={TechTrends},
  volume={65},
  number={5},
  pages={796--807},
  year={2021},
  publisher={Springer}
}

@inproceedings{wu2025enhancing,
  title={Enhancing Elementary Teachers’ AI-TPACK through a Professional Development Workshop},
  author={Wu, Ying-Tien and Wang, Li-Jen and Shih, Tsun-Hui},
  booktitle={Proceedings of the 19th International Conference of the Learning Sciences-ICLS 2025, pp. 2554-2556},
  year={2025},
  organization={International Society of the Learning Sciences}
}

@article{rush2016impact,
  title={The impact of item-writing flaws and item complexity on examination item difficulty and discrimination value},
  author={Rush, Bonnie R and Rankin, David C and White, Brad J},
  journal={BMC medical education},
  volume={16},
  number={1},
  pages={250},
  year={2016},
  publisher={Springer}
}

@article{sabatini2020engineering,
  title={Engineering a twenty-first century reading comprehension assessment system utilizing scenario-based assessment techniques},
  author={Sabatini, John and O’Reilly, Tenaha and Weeks, Jonathan and Wang, Zuowei},
  journal={International Journal of Testing},
  volume={20},
  number={1},
  pages={1--23},
  year={2020},
  publisher={Taylor \& Francis}
}

@incollection{darzhinova2025technology,
  title={Technology-enhanced Scenario-based Reading Assessment of Pre-service English Teachers},
  author={Darzhinova, Liubov},
  booktitle={The Routledge Handbook of the Sociopolitical Context of Language Learning},
  pages={470--489},
  year={2025},
  publisher={Routledge}
}

@article{laupichler2023delphi,
  title={Delphi study for the development and preliminary validation of an item set for the assessment of non-experts' AI literacy},
  author={Laupichler, Matthias Carl and Aster, Alexandra and Raupach, Tobias},
  journal={Computers and Education: Artificial Intelligence},
  volume={4},
  pages={100126},
  year={2023},
  publisher={Elsevier}
}

@article{laupichler2023development,
  title={Development of the “Scale for the assessment of non-experts’ AI literacy”--An exploratory factor analysis},
  author={Laupichler, Matthias Carl and Aster, Alexandra and Haverkamp, Nicolas and Raupach, Tobias},
  journal={Computers in Human Behavior Reports},
  volume={12},
  pages={100338},
  year={2023},
  publisher={Elsevier}
}

@article{zhang2025developing,
  title={Developing and validating the artificial intelligence literacy concept inventory: An instrument to assess artificial intelligence literacy among middle school students},
  author={Zhang, Helen and Perry, Anthony and Lee, Irene},
  journal={International Journal of Artificial Intelligence in Education},
  volume={35},
  number={1},
  pages={398--438},
  year={2025},
  publisher={Springer}
}

@article{iqbal2025teaching,
  title={Teaching AI Awareness in Middle School Classrooms: Design, Implementation and Evaluation of Two Education Modules on Algorithmic Bias and Filter Bubbles},
  author={Iqbal, Mehtab and Singh, Khushbu and Khan, Sushmita and Osho, Oluwafemi and Sidnam-Mauch, Emily and Bannister, Nicole and Caine, Kelly and Knijnenburg, Bart},
  journal={Computers and Education: Artificial Intelligence},
  pages={100425},
  year={2025},
  publisher={Elsevier}
}

@inproceedings{black2024framework,
  title={A framework for approaching AI education in educator preparation programs},
  author={Black, Nancye Blair and George, Stacy and Eguchi, Amy and Dempsey, J Camille and Langran, Elizabeth and Fraga, Lucretia and Brunvand, Stein and Howard, Nicol},
  booktitle={Proceedings of the AAAI Conference on Artificial Intelligence},
  volume={38},
  number={21},
  pages={23069--23077},
  year={2024}
}

@inproceedings{long2020ai,
  title={What is AI literacy? Competencies and design considerations},
  author={Long, Duri and Magerko, Brian},
  booktitle={Proceedings of the 2020 CHI conference on human factors in computing systems},
  pages={1--16},
  year={2020}
}

@misc{chiu2025responsible,
  title={Responsible digital citizen: building AI ethics awareness across subjects},
  author={Chiu, Thomas KF},
  journal={Interactive Learning Environments},
  volume={33},
  number={4},
  pages={2759--2761},
  year={2025},
  publisher={Taylor \& Francis}
}

@misc{unesco2021ai,
  author       = {{UNESCO}},
  title        = {Artificial Intelligence and Education: Guidance for Policy-makers},
  year         = {2021},
  howpublished = {\url{https://unesdoc.unesco.org/ark:/48223/pf0000376709}},
  note         = {Policy Report}
}

@misc{unesco2023genai,
  author       = {{UNESCO}},
  title        = {Guidance for Generative AI in Education and Research},
  year         = {2023},
  howpublished = {\url{https://www.unesco.org/en/articles/guidance-generative-ai-education-and-research}},
  note         = {Ethical Guidance Report}
}

@misc{whitehouse2025ai,
  author       = {{The White House}},
  title        = {Executive Order on AI Education},
  year         = {2025},
  month        = {April},
  day          = {23},
  howpublished = {\url{https://www.whitehouse.gov/presidential-actions/2025/04/advancing-artificial-intelligence-education-for-american-youth/}},
  note         = {Executive Order}
}

@misc{eu2021digital,
  author       = {{European Commission}},
  title        = {Digital Education Action Plan (2021–2027)},
  year         = {2021},
  howpublished = {\url{https://education.ec.europa.eu/focus-topics/digital-education/action-plan}},
  note         = {Policy Framework}
}

@misc{china2025edustrategy,
  author       = {{Ministry of Education of the People’s Republic of China}},
  title        = {Education Modernization Plan 2024–2035},
  year         = {2025},
  url = {https://www.gov.cn/zhengce/202501/content\_7000579.htm}
}

@misc{naer108curriculum,
  author       = {{National Academy for Educational Research}},
  title        = {108 Curriculum Guidelines},
  howpublished = {\url{https://www.naer.edu.tw/eng/PageSyllabus?fid=148}},
  year      = {2024}
}

@online{figueiredo2025ai_literacy_programs,
  author      = {Figueiredo, T{\'a}nia},
  title       = {AI Literacy Programs in Europe},
  year        = {2025},
  month       = {5},
  day         = {23},
  organization= {Future of Life Institute},
  howpublished= {\url{https://artificialintelligenceact.eu/ai-literacy-programs/}},
  note        = {EU Artificial Intelligence Act website},
  urldate     = {2025-09-22}
}

@online{iste_key_initiatives,
  author       = {{International Society for Technology in Education (ISTE)}},
  title        = {Key Initiatives},
  howpublished = {\url{https://iste.org/key-initiatives}},
  note         = {Accessed: 2025-09-22}
}

@online{unesco_ai_cft_2024,
  author       = {Miao, Fengchun and Cukurova, Mutlu},
  title        = {AI Competency Framework for Teachers},
  year         = {2024},
  organization = {UNESCO},
  url          = {https://www.unesco.org/en/articles/ai-competency-framework-teachers},
  urldate      = {2025-09-22},
  note         = {Published 8 Aug 2024; last updated 18 Aug 2025}
}

@article{kuder1937theory,
  title={The theory of the estimation of test reliability},
  author={Kuder, G Frederic and Richardson, Marion W},
  journal={Psychometrika},
  volume={2},
  number={3},
  pages={151--160},
  year={1937},
  publisher={Springer}
}

@article{boone2016rasch,
  title={Rasch analysis for instrument development: why, when, and how?},
  author={Boone, William J},
  journal={CBE—Life Sciences Education},
  volume={15},
  number={4},
  pages={rm4},
  year={2016},
  publisher={American Society for Cell Biology}
}

@article{topal2025artificial,
  title={Artificial intelligence literacy scale: A study of reliability and validity in Turkish university students},
  author={Topal, Arzu Deveci and G{\"o}k{\c{c}}e, Asiye Toker and Eren, Canan Dilek and Ge{\c{c}}er, Aynur Kolburan},
  journal={Journal of Learning and Teaching in Digital Age},
  volume={10},
  number={1},
  pages={58--67},
  year={2025},
  publisher={Mehmet Akif OCAK}
}

@article{kaiser1974index,
  title={An index of factorial simplicity},
  author={Kaiser, Henry F},
  journal={psychometrika},
  volume={39},
  number={1},
  pages={31--36},
  year={1974},
  publisher={Springer}
}

@article{bewersdorff2025ai,
  title={How AI literacy correlates with affective, behavioral, cognitive and contextual variables: A systematic review},
  author={Bewersdorff, Arne and Nerdel, Claudia and Zhai, Xiaoming},
  journal={Computers and Education: Artificial Intelligence},
  pages={100493},
  year={2025},
  publisher={Elsevier}
}

@misc{google-ai-educators,
  title        = {AI for Educators – Grow with Google},
  author       = {{Google}},
  howpublished = {\url{https://grow.google/ai-for-educators/}},
  note         = {Accessed: 2025-12-09}
}

\end{document}